\documentclass{pasj01}

\Received{2015/11/26}
\Accepted{2016/04/21}
\Published{}
\SetRunningHead{High-frequency excess in radio spectrum of NGC~985}{Doi \& Inoue}

\usepackage{times}
\usepackage{natbib}
\begin{document}

\title{High-frequency excess in the radio continuum spectrum of the type-1 Seyfert galaxy NGC~985}
\author{
Akihiro \textsc{Doi}\altaffilmark{1,2} and  
Yoshiyuki \textsc{Inoue}\altaffilmark{1} 
}
\altaffiltext{1}{The Institute of Space and Astronautical Science, Japan Aerospace Exploration Agency, 3-1-1 Yoshinodai, Chuou-ku, Sagamihara, Kanagawa 252-5210}
\altaffiltext{2}{Department of Space and Astronautical Science, The Graduate University for Advanced Studies (SOKENDAI), 3-1-1 Yoshinodai, Chuou-ku, Sagamihara, Kanagawa 252-5210}


\maketitle

\begin{abstract}

The Seyfert galaxy NGC~985 is known to show a high-frequency excess in its radio continuum spectrum in a milli-Jansky level on the basis of previous observations at 1.4--15~GHz; a steep spectrum at low frequencies (a spectral index of $\alpha=-1.10 \pm 0.03$) changes at $\sim10$~GHz into an inverted spectrum at higher frequencies ($\alpha=+0.86 \pm 0.09$).  We conduct new observations at 15--43~GHz using the Very Large Array and at 100~GHz using the Nobeyama Millimeter Array.  As a result, the high-frequency excess continuing at even higher radio frequencies up to 43~GHz has been confirmed.  The non-detection at 100~GHz was not so strong constraint, and therefore the spectral behavior above 43~GHz remains unclear.   The astrometric position of the high-frequency excess component coincides with the optical position of the Seyfert nucleus and the low-frequency radio position to an accuracy of 0.1~arcsec, corresponding to $\sim80$~pc; the radio source size is constrained to be $<0.02$~arcsec, corresponding to $<16$~pc.  We discuss the physical origin of the observed high-frequency excess component.  
Dust emission at the Rayleigh-Jeans regime, free--free emission from X-ray radiating high-temperature plasma, free--free emission from the ensemble of broad-line region clouds, or thermal synchrotron from hot accretion flow cannot be responsible for the observed radio flux.  Compact jets under synchrotron self-absorption may be unlikely in terms of observed time scales.  
Alternatively, we cannot rule out the hypotheses of synchrotron jets free--free absorbed by a circumnuclear photo-ionized region, and self-absorbed nonthermal synchrotron from disk corona, as the origin of the high-frequency excess component.

\end{abstract}

\section{Introduction}\label{section:introduction}

Active galactic nuclei~(AGNs) are believed to be powered by gravitational energy released from accreting matter onto supermassive black holes at the center of galaxies.  The electromagnetic spectra of AGNs are observed in broadband ranging from radio to $\gamma$-ray, however, the AGN spectra are possibly contaminated by the components of stellar processes related to star-forming activity in their host galaxies.  
In the radio regime at lower than $\sim10$~GHz, optically-thin synchrotron emissions from AGN jets and/or supernova remnants dominate the spectral energy distributions~(SEDs) and are generally thought to exhibit a steep spectrum with a spectral index of $\alpha<-0.5$, where $S_\nu \propto \nu^{\alpha}$ and $S_\nu$ is the flux density at the frequency $\nu$.  However, AGNs show a wide variety of radio spectral shapes.  Blazars generally show flat spectra ($\alpha\sim0$) with significant flux variations, as a result of dominance of radio cores by relativistic beaming effect due to innermost fast jets aligned close to our line of sight \citep{Blandford:1979}.  Compact strong radio sources with no significant variation frequently exhibit convex spectra peaking at $\sim100$~MHz (compact steep spectrum: CSS) and $\sim1$~GHz (Gigahertz-peaked spectrum: GPS), which are caused by absorption of low-frequency radio on small-scaled radio galaxies that are young or reactivated recently (\citealt{ODea:1998}, and references therein).   Nearby low-luminosity AGNs are frequently detected as weak radio sources, which show flat or slightly inverted spectra for a spatially-extracted nuclear component \citep{Nagar:2001,Ulvestad:2001,Anderson:2004,Doi:2005a}; their origin is still unclear, although a compact jet base powered by radiatively inefficient accretion flows~(RIAF) at a low accretion rate is the most promising candidate \citep{Falcke:1996,Ulvestad:2001,Merloni:2003,Markoff:2008,de-Gasperin:2011,Plotkin:2012}.  Such  component of flat or slightly inverted spectra sometimes dominate at millimeter bands ($\sim100$~GHz) rather than the steep spectrum synchrotron emission, free--free emission, and dust emission associated with the host galaxies \citep{Doi:2005,Doi:2011}.

\citet{Antonucci:1988} and \citet{Barvainis:1996} reported the discovery of ``high frequency excess (HFE)'' in radio continuum spectra of several radio-quiet quasars/luminous Seyferts on the basis of early VLA observations at 1.5--14.9~GHz.  The HFEs show a steep spectrum at low frequencies but flatter or inverted at higher frequencies ($\gtrsim10$~GHz).  The physical origin of the high-frequency component is unclear.  These luminous AGNs might not be the case applicable to the RIAF for low accretion rates.  \citet{Antonucci:1988} discussed the physical validity of self-absorbed synchrotron from a compact AGN radio core and free--free emission related to AGN and stellar activities.  On the basis of the coexistence of radio variability and flat or inverted spectrum as well as a VLBI detection, \citet{Barvainis:1996} suggested that radio cores as scaled-down versions of those seen in radio loud quasars are more likely in these radio-quiet quasars.     
\citet{Behar:2015} also found millimeter radio excesses in several radio quiet AGNs, and suggest accretion disk coronal emissions as its origin (see also \citealt{Laor:2008}).  \citet{Inoue:2014} predict a synchrotron luminosity detectable in the submillimeter regime if nonthermal electrons in addition to thermal ones are present in a magnetized disk corona in a nearby Seyfert galaxy.

NGC~985 (Mrk~1048), a luminous Seyfert type-1 galaxy located at a redshift of 0.0427, was one of the sample of \citet{Antonucci:1988} and \citet{Barvainis:1996} and showed one of the most prominent signature of the HFE: a steep spectrum at 1.4--8.4~GHz and an inverted spectrum at 8.4--15~GHz.  
The luminosity of NGC~985 is $L_\mathrm{IR} =1.8 \times 10^{11} L_\odot$, which puts this object in the luminous infrared galaxy group.  
This is a peculiar galaxy with a prominent ring-shaped zone at several kiloparsecs from the nucleus \citep{de-Vaucouleurs:1975}, suggesting that the galaxy is undergoing a merging process.  The galaxy contains a double nucleus \citep[a Seyfert nucleus and a non-AGN;][and references therein]{Perez-Garcia:1996}; the nuclei are separated by 3~arcsec \citep[2.5~kpc of projected distance;][]{Appleton:2002}, which indicates that the intruder galaxy responsible for the formation of the ring is sinking in the nuclear potential of the primary galaxy \citep{Perez-Garcia:1996}.    
An ionized absorber has been observed in the X-ray spectra of NGC~985, indicating the presence of ionized outflows with velocities of $\sim200$--500~km/s \citep{Krongold:2005,Parker:2014}, as well as a hard X-ray component suggesting inverse-Compton radiation from corona on an accretion disk of the Seyfert nucleus.  
Studies including other radio observations have been made by \citet{Ulvestad:1984,Appleton:2002}, which mainly mentioned a low-frequency ($\leqq8.5$~GHz) radio structure on the basis of VLA observations at higher angular resolutions.  These results will be incorporated into our study for the spectrum of NGC~985 at the radio-to-FIR regime.

The present paper reports finding of the millimeter excess in the radio continuum spectrum of NGC~985 nucleus, on the basis of new radio observations at higher frequencies in combination with archival data and published results.     
In Section~\ref{section:dataanddataanalysis}, we describe the data, data reduction, and analyses of the radio data.  The results are presented in Section~\ref{section:result}, and their implications are discussed in Section~\ref{section:discussion}.  Finally, we provide summary and conclusions in Section~\ref{section:conclusion}.    
Throughout this paper, we use Lambda cold dark matter ($\Lambda$CDM) cosmology with $H_0=70.5$~km~s$^{-1}$~Mpc$^{-1}$, $\Omega_\mathrm{M}=0.27$, and $\Omega_\mathrm{\Lambda}=0.73$.  The redshift is $z=0.043143 \pm 0.000073$~\citep{Fisher:1995}; the luminosity distance is $186$~Mpc, and the angular-size distance is $171$~Mpc; 1~arcsecond corresponds to a projected linear scale of $831$~pc at the distance to NGC~985.

\section{Data and Data Analysis}\label{section:dataanddataanalysis}
\subsection{New VLA observations}\label{section:newVLAobservation}  
We conducted new VLA observations toward NGC~985 at 22~GHz on 2001 September~28 with the CnD-configuration (AD456) and at quasi-simultaneous 8.46--43.3~GHz on 2003 December~24~with the B-configuration (AD489).  The 15, 22, and 43-GHz observations were carried out by fast switching of antenna pointing between the target source and a nearby calibrator to reduce atmospheric phase fluctuations for long baselines.  The 8.46~GHz observation was carried out in the standard manner.  

We reduced the data in the standard manners using the Astronomical Image Processing System \citep[AIPS; ][]{Greisen:2003}, developed at the National Radio Astronomy Observatory~(NRAO).  The calibrator 3C~48 served a flux scaling factor at each frequency.  We followed the guidelines for accurate flux density bootstrapping \citep{Perley:2003}, including correcting the dependence of the gain curve on elevation for the antennas and atmospheric opacity and by using the source structure models in self-calibrations.  The bootstrap accuracy should be 1\%--2\% at 20, 6, and 3.6~cm, and perhaps 3\%--5\% at 2, 1.3, and 0.7~cm \citep{Perley:2003}.  In the present study, for scaling factors, we assumed uncertainties of 3\% and 5\% at 8.46~GHz and 14.9--43.3~GHz, respectively.  All target images were synthesized from calibrated visibilities with natural weighting and deconvolved with the CLEAN algorithm using the AIPS task IMAGR.  

\subsection{VLA archival data}\label{section:VLAarchivaldata}  

We retrieved the VLA archival data AN114, which was observed with the VLA-A at 8.46~GHz.  
This data set provides an image at the highest angular resolution ($\timeform{0.35''}\times\timeform{0.25''}$) with the best sensitivity of $0.04$~mJy~beam$^{-1}$ in this study.  This allows us to determine a precise astrometric position of the object.  

We also retrieved the VLA archival data AA48 (VLA-C) and AB489 (VLA-D/A), which had been used in \citet{Barvainis:1996}; they pointed out the presence of high-frequency excess~(HEF) from these data at 1.49--14.9~GHz.  AA48 and AB489 originally include observations at 22.5~GHz as well as 1.49--14.9~GHz, however, results at 22.5~GHz were not reported in \citet{Barvainis:1996}.  In the present study, the 22-GHz data are analyzed.  According to imaging procedure in \citet{Barvainis:1996}, AB489 data with the VLA-D/A hybrid array was imaged from visibilities tapered to 100~k$\lambda$ resulting in $\sim\timeform{4''}$ resolution.

We reduced these archival data in the same manners using the AIPS, as described in Section~\ref{section:newVLAobservation}.  AN114 data and AA48/AB489 data showed the positive and non-detections, respectively.    

\citet{Ulvestad:1984} reported the result of VLA observation with the A-configuration at 4.89~GHz as a total flux density of $2.5 \pm 0.6$~mJy from heavily tapered visibility data and $\sim1$~mJy in the untapered map for this galaxy.  For the present study, we retrieved the image data of \citet{Ulvestad:1984} from the NASA/IPAC Extragalactic Database~(NED) archive, and then, analyzed it and found a compact component of $1.2 \pm 0.3$~mJy for the untapered case.

\subsection{NMA observation}\label{section:NMAobservation} 
 
We observed NGC~985 using the Nobeyama Millimeter Array~(NMA) on 2003~April~26, May~14, 24, and 25.  We used the D-configuration that is the most compact array configuration, resulting in half-power beam widths (HPBWs) of synthesized beams of $\sim 7 \arcsec$.  Visibility data were obtained with double-sided receiving system at centre frequencies of 89.725 and 101.725~GHz, where each band has a bandwidth of 1~GHz, i.e., 2~GHz in total.  The Ultra Wide-Band Correlator \citep[UWBC;][]{Okumura:2000} was used.  Typical system noise temperature, $T_\mathrm{sys}$, is about 150~K.  We made scans of B0238$-$084~(NGC~1052) every 25~minutes for gain calibration.  Bandpass calibrators of 3C~279 or 3C~454.3 were observed once a day.  The data were reduced using the UVPROC-II package, developed at the Nobeyama Radio Observatory~(NRO).  Visibilities of both the sidebands were combined with the same weight, which resulted in an averaged centre frequency of 95.725~GHz ($\lambda3.1$~mm).  Each daily image was individually processed in natural weighting, and then deconvolved using the AIPS.  Furthermore, the visibilities of all the days were combined with a weight of $T_\mathrm{sys}^{-2}$, resulting in an aggregated observing time of 12.6~hours, and then imaged.  Flux scales of the gain calibrator were derived with the uncertainty to 10~per~cent by relative comparisons to flux-known calibrators like the Uranus, which were scanned quasi-simultaneously at almost the same elevations.  Consequently, no significant emission appeared above three-times the r.m.s.~of image noise ($1\sigma = 1.52$~mJy~beam$^{-1}$).

\section{Result}\label{section:result}

\subsection{Flux densities and radio morphology}\label{section:fluxdensity}

The new results are listed in Table~\ref{table:observation} and plotted in Figure~\ref{figure:spectrum}, which also include data from the literature.  In all detected cases, a single emission component was found.  Its flux density and deconvolved size were measured using the AIPS task JMFIT, Gaussian-profile model fitting.  The error of the flux measurements was determined by root-sum-squares of image noise, the fitting error, and the flux scaling uncertainty.  In non-detections, we defined three times the image r.m.s.~noise ($3\sigma_\mathrm{rms}$) as the upper limit of flux density.  The source was apparently unresolved in any detected cases.  The radio source size was constrained to be $<\timeform{0.02''}$, corresponding to $<16$~pc; the VLA observation AN114 provided an image with the highest quality and the upper limit of deconvolved size with a signal-to-noise ratio of $18.7$ and a major-axis beam size of $\timeform{0.35''}$.   

Radio photometry using interferometry is sometimes affected by resolution effect if a radio structure is significantly larger than fringe spacings of its shortest baselines.  
At $\sim1.4$~GHz, the NVSS flux density (VLA-D at a $\timeform{45''}$ resolution) is significantly larger than the results of VLA-C and VLA-D/A hybrid (tapered to 100~k$\lambda$ resulting in $\sim\timeform{4''}$ at all frequencies) by \citet{Barvainis:1996} (Figure~\ref{figure:spectrum}); this is clear evidence of resolution effect.    
The radio morphology of NGC~985, which has been presented by \citet{Appleton:2002} on the basis of VLA-D observations at 4.9 and 8.4~GHz, exhibited a twin-peaks (separated by $\sim\timeform{20''}$) with a brighter component associated with the position of the Seyfert nucleus and a second one which lies to the west just inside the outer ring of the host galaxy, the latter of which is presumably related to star forming activity.  
\citet{Ulvestad:1984} reported the result of VLA-A observation at 4.89~GHz toward the Seyfert nucleus as a total flux density of $2.5 \pm 0.6$~mJy from heavily tapered visibility data while $\sim1$~mJy in the untapered map ($1.2 \pm 0.3$~mJy by our re-analysis; Section~\ref{section:VLAarchivaldata}); therefore, the authors pointed out that the bulk of the radio emission toward the AGN is probably from an extended, low surface brightness component.  The VLA-D observation at 4.9~GHz showing 3.38~mJy for the nucleus supports this picture \citep{Appleton:2002}.  
Thus, a substantial fraction of the brighter emission itself originates in low-brightness component.  Similarly, the discrepancy between our VLA-A and VLA-B results at 8.46~GHz (Table~\ref{table:observation}) is possibly attributed to the resolution effect on such a radio structure.  Thus, as in the lower-frequency emission ($\lesssim10$~GHz), it is unclear whether the physical origin is AGN jets or star forming activity, which is discussed later in Section~\ref{section:discussion}.  On the other hand, we have no evidence of resolving out for the higher-frequency emission.   

The contribution from the secondary nucleus, which is a non-AGN and separated by $\sim\timeform{3''}$, is negligible for the radio photometry toward the Seyfert nucleus.  No signature of the radio emission appeared at the putative location of the secondary nucleus in the VLA-A image with the highest sensitivity ($0.04$~mJy~beam$^{-1}$ at 8.46~GHz) in the present study, even in a heavily tapered map.

\subsection{Radio Spectrum}\label{section:radiospectrum}

\begin{figure*}
\begin{center}
\FigureFile(1\linewidth, ){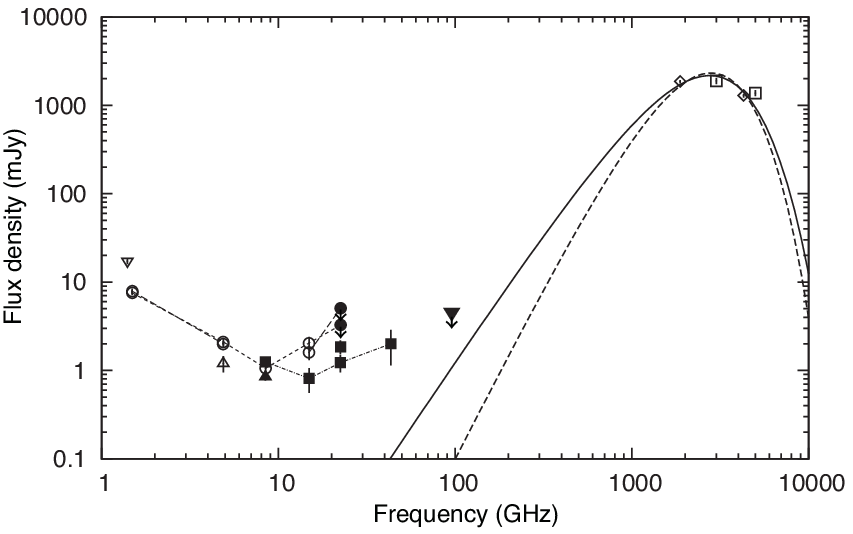}
\end{center}
\caption{Radio-to-FIR spectrum of NGC~985 nucleus.  Open and Filled symbols are data published by other authors and data newly-presented in the present study, respectively.  Non-detections are indicated by downward arrows.  Lines connecting symbols indicate quasi-simultaneous observations.  Filled Squares: our VLA observations with VLA-CnD at 22.5~GHz (AD456) and VLA-B at 8.46--43.3~GHz (AD489).   Lower-peaked filled triangle: NMA observation at 95.7~GHz as non-detection.  Upper-peaked triangles: VLA-A observations; filled symbol at 4.89~GHz is data of \citet{Ulvestad:1984} and reanalyzed in the present study; open symbol at 8.46~GHz is from  archival data AN114.  Open circles: VLA-C (AA48) and VLA-D/A hybrid (AB489, tapered to 100~k$\lambda$ resulting in $\sim\timeform{4''}$ at all frequencies) observations reported by \citet{Barvainis:1996}; 22.5-GHz data (non-detections) are newly reported in the present paper.  Lower-peaked open triangle: NVSS result \citep{Condon:1998}.  Open diamonds: {\it Herschel} PACS at 70~$\mu$m and 160~$\mu$m toward the nucleus \citep{Melendez:2014}.  Open squares: {\it IRAS} Faint Source Catalogue, version 2.0 \citep{Moshir:1990} at 60~$\mu$m and 100~$\mu$m.  Solid and dashed curves: dust model spectra for cases of the emissivity $\beta=1$ (33.7~K) and $\beta=2$ (27.1~K), respectively.  
} 
\label{figure:spectrum}
\end{figure*}

The resultant radio spectrum (Figure~\ref{figure:spectrum}) consists of two components: (1)~a steep spectrum from $\sim1$~GHz to $\sim10$~GHz ($\alpha=-1.10 \pm 0.03$, where $S_\nu \propto \nu^{\alpha}$), and (2)~an inverted spectrum from $\sim10$~GHz at least up to $\gtrsim40$~GHz ($\alpha=+0.86 \pm 0.09$); we used the data of AA48, AB486, and AD489, which were obtained at multi-frequency quasi-simultaneously, for determining these spectral indices.       
The HFE previously discovered at 8.46--14.9~GHz by \citet{Barvainis:1996} (open squares in Figure~\ref{figure:spectrum}); two 22.5~GHz observations in 1985 and 1990, which had not been published although observed quasi-simultaneously with the lower frequencies, showed upper limits that are not so strong constraint to rule out the continuity of the HFE.  Our observation on 2001 at 22.5~GHz showed a detection and provided evidence for the presence of the HFE at a higher frequency.  On 2003, our NMA observation resulted in a non-detection, which provides an insufficient constraint to probe a potential peak on spectrum.  Finally, our VLA observations at quasi-simultaneous multi-frequency showed a clear evidence of the HFE up to 43.3~GHz.   

The transition frequencies between a steep and an inverted spectra ($\sim10$~GHz) on 1990 and 2003 seem to be different.

\subsection{Radio Source Position}\label{section:radioposition}

The optical position of the Seyfert nucleus is R.A.$=\timeform{02h34m37.840s}$, Decl.$= \timeform{-08D47'16.07''}$ (an error of $\sim\timeform{0.1''}$) on the basis of the Sloan Digital Sky Survey Data Release~9 \citep[SDSS~DR9;][]{Ahn:2012}.  On the other hand, the highest resolution radio image at 8.46~GHz showed R.A.$=\timeform{02h34m37.84024s} \pm \timeform{0.00040s}$, Decl.$= \timeform{-08D47'16.0877''}\pm \timeform{0.0083''}$, which is consistent with the SDSS position.   The high frequency radio images also showed positions consistent with the SDSS DR9: 
R.A.$=\timeform{02h34m37.83962s} \pm \timeform{0.00134s}$, 
Decl.$= \timeform{-08D47'16.0485''}\pm \timeform{0.0262''}$ at 43.3~GHz and 
R.A.$=\timeform{02h34m37.84016s} \pm \timeform{0.00197s}$, 
Decl.$= \timeform{-08D47'16.0509''}\pm \timeform{0.0305''}$ at 22.5~GHz in VLA-B.    
Thus, the HFE component is presumably associated with the Seyfert nucleus to an accuracy of $\sim\timeform{0.1''}$.

\begin{table*}
\caption{Results of observations.\label{table:observation}}
\begin{center}
\begin{footnotesize}
\begin{tabular}{lcccccccr}
\hline
\hline
Obs. date & VLA Obs. ID & Ref. & Array & $\nu$ & $S_\nu$   & $\sigma_\mathrm{rms}$ & $\theta_\mathrm{maj} \times \theta_\mathrm{min}$   & $\phi_\mathrm{PA}$ \\
 &  &  &  & (GHz) & (mJy)   & (mJy beam$^{-1}$) & (arcsec$\times$arcsec)   & (deg) \\
(1) & (2) & (3) & (4) & (5) & (6)   & (7) & (8)   & (9) \\
\hline
1982 May 30 & (N/A) & a & VLA-A & 4.89 & $ 1.2 \pm 0.25 $ & 0.10  & $ 5.8  \times 4.4 $ & $ -18.7 $ \\
1985 Jul 28 & AA48 & b & VLA-C & 22.5 & $<5.0$   & 1.68 & $ 1.5 \times 1.1 $ & $ -20.0 $ \\
 &  & c &  & 14.9 & $ 1.6 \pm 0.29 $ & \ldots & $ \sim 4 \times 4 $ &  \ldots \\
1986 Nov 08 &  & c &  & 4.89 & $ 2.0 \pm 0.15 $ & \ldots & $ \sim 4 \times 4 $ &  \ldots \\
 &  & c &  & 1.49 & $ 7.9 \pm 0.53 $ & \ldots & $ \sim 4 \times 4 $ &  \ldots \\
1990 May 23 & AB486 & b & VLA-D/A & 22.5 & $<3.3$   & 1.09 & $ 5.6 \times 3.9 $ & $ 59.0 $ \\
 &  & c &  & 14.9 & $ 2.0  \pm 0.35 $ & \ldots & $ \sim 4 \times 4 $ &  \ldots \\
 &  & c &  & 8.48 & $ 1.1  \pm 0.24 $ & \ldots & $ \sim 4 \times 4 $ &  \ldots \\
 &  & c &  & 4.89 & $ 2.1  \pm 0.18 $ & \ldots & $ \sim 4 \times 4 $ &  \ldots \\
 &  & c &  & 1.49 & $ 7.6  \pm 0.44 $ & \ldots & $ \sim 4 \times 4 $ &  \ldots \\
1993 Sep 20 & (NVSS) &  d &  VLA-D & 1.4 & $ 17.3 \pm 1.3 $ & 0.45 & $ 45 \times 45 $ & $ 0.0 $ \\
2001 Sep 28 & AD456 & b & VLA-CnD & 22.5 & $ 1.9 \pm 0.3 $ & 0.15 & $ 3.8 \times 1.7 $ & $ 72.3 $ \\
2003 Apr 03--May 25 & \ldots & b & NMA-D & 95.7 & $<4.6$   & 1.52 & $ 8.1 \times 6.1 $ & $ -12.8 $ \\
2003 Jun 19 & AN114 & b & VLA-A & 8.46 & $ 0.84 \pm 0.08 $ & 0.04 & $ 0.35 \times 0.25 $ & $ 12.0 $ \\
2003 Dec 24 & AD489 & b & VLA-B & 43.3 & $ 2.0 \pm 0.9 $ & 0.48 & $ 0.35 \times 0.14 $ & $ -30.2 $ \\
 &  & b &  & 22.5 & $ 1.2 \pm 0.3 $ & 0.13 & $ 0.56 \times 0.31 $ & $ -35.2 $ \\
 &  & b &  & 14.9 & $ 0.81 \pm 0.25 $ & 0.25 & $ 0.88 \times 0.44 $ & $ -39.3 $ \\
 &  & b &  & 8.46 & $ 1.3 \pm 0.1 $ & 0.10 & $ 1.7 \times 0.8 $ & $ -41.2 $ \\
\hline
\end{tabular}
\end{footnotesize}
\end{center}
\begin{flushleft}
\begin{footnotesize}
Col.~(1) observation date; 
Col.~(2) VLA observation ID; Col.~(3) Reference. a: \citealt{Ulvestad:1984}, b: the present study, c: \citealt{Barvainis:1996}, d: \citealt{Condon:1998};   
Col.~(4) array configuration; 
Col.~(5) center frequency; 
Col.~(6) total flux density; 
Col.~(7) image rms~noise on blank sky; 
Cols.~(8)--(9) synthesized beam sizes in major axis, minor axis, and position angle of major axis, respectively.  
\end{footnotesize}
\end{flushleft}
\end{table*}

\section{Discussion}\label{section:discussion}

The HFE has been confirmed in the radio spectrum toward the Seyfert nucleus of NGC~985 by our new VLA observations up to 43.3~GHz ($\lambda7$mm); a steep spectrum ($\alpha=-1.1$) changes into an inverted spectrum ($\alpha\approx+0.9$) at $\sim10$~GHz (Section~\ref{section:radiospectrum}).  The non-detection in our NMA observation at 95.7~GHz ($\lambda3$mm) suggests a peak-out frequency between $43.3$ and $95.7$~GHz.  However, this upper limit was not so conclusive.  Therefore, the spectral peak-out at a higher frequency cannot be ruled out on the basis of the present study.  The sensitivities of the Atacama Large Millimeter/submillimeter Array~(ALMA) have potential to reveal the detailed spectral profile in the near future, which leads to distinguish physical origins of the HFE.  In the present paper, we provide basic discussions of the physical origins with limited information at this time.           

Before the discussion for the HFE, the physical origin of the low-frequency spectral component is addressed.  
The steep spectrum with the index $\alpha=1.1$ is the indication of optically-thin synchrotron radiation originating in starburst and/or AGN activities.  
The standard FIR/radio ratio $q$-value \citep[][and references therein]{Condon:1992} is 2.37, indicating that the low-frequency radio emission is explainable by star-forming activity, according to $2.34 \pm 0.19$ for the {\it IRAS} Bright Galaxy Sample starbursts \citep{Condon:1991a}.  Here we used the radio flux density 7.9~mJy at 1.49~GHz \citep{Barvainis:1996} and the {\it IRAS} flux densities 1.893~Jy and 1.381~Jy at 100~$\mu$m and 60~$\mu$m, respectively.  A significant contribution from the extended, low surface brightness component that was inferred at $\lesssim10$~GHz (Section~\ref{section:fluxdensity}) agrees with this picture as stellar process in the host galaxy; the AGN jet contribution would not be necessary.

\subsection{Dust Emission: insufficient for HFE}\label{section:dust}

The potential contribution from dust into the high-frequency radio regime can be restricted by an extrapolation from far-infrared~(FIR) observations using a gray body model spectrum; $S_\nu \propto \nu^{\beta} B(T_\mathrm{d}$), where $B(T_\mathrm{d})$ is the blackbody spectrum at temperature $T_\mathrm{d}$ \citep[e.g.,][and references therein]{Temi:2004}.  The dust emissivity $\beta$ affects the slope in the Rayleigh-Jeans region: small values produce a flatter slope and the maximal estimation for lower frequencies.  Almost all observations of external galaxies show $\beta \sim 1.6$--2 \citep[][and references therein]{Temi:2004}.  Even if we adopt an extreme value of $\beta=1$, $<0.10$~mJy at 43.3~GHz is obtained based on the estimation using the {\it Herschel} PACS observations \citep{Melendez:2014} and {\it IRAS} Faint Source Catalogue, version~2.0 \citep{Moshir:1990} (Figure~\ref{figure:spectrum}).  Thus, dust cannot be responsible for the observed HFE component.

\subsection{Free--Free Emission: insufficient for HFE}\label{section:FFE}
An ionized region radiates free--free emission~(FFE) from radio to X-ray regimes.  
For the NGC~985 nucleus, the broadband X-ray (0.1--100~keV) continuum from the NGC~985 nucleus can be parameterized by a photon spectral power law ($\Gamma \sim 1.4$) and a high-energy cutoff ($E_\mathrm{c} \sim70$~keV) with a flux of $f(2$--10~keV)$ = 1.49 \times 10^{-11}$~erg~s$^{-1}$~cm$^{-2}$ after the correction of  the Galactic absorption (\citealt{Krongold:2005}; $6.2 \times 10^{43}$~ergs~s$^{-1}$), which is consistent with another estimation: a 2--10-keV luminosity of $4.2 \times 10^{43}$~ergs~s$^{-1}$ after correcting for both Galactic and intrinsic absorption \citep{Parker:2014}.  
We assume all the X-ray emission originates via the FFE process from ionized gas with an electron temperature of $\sim1.6\times10^8$~K ($E_\mathrm{c} \sim70$~keV), although inverse-Compton radiation and/or synchrotron emission may also contribute the observed X-ray in addition to FFE.  
Since the ratio of Gaunt factors at radio and X-ray can be predicted, we can simply diagnose whether observed radio and X-ray flux densities are the same origin or not, as demonstrated by \citet{Laor:2008,Steenbrugge:2011}, by using a more generally expressed equation:    
\begin{equation}
\frac{S_{\nu_1}}{S_{\nu_2}} = e^{-h(\nu_1 - \nu_2)/ k_\mathrm{B} T_\mathrm{e}} \frac{\bar{g_\mathrm{ff}}(\nu_1, T_\mathrm{e})}{\bar{g_\mathrm{ff}}(\nu_2, T_\mathrm{e})},      
\label{equation:radio2X-ray}
\end{equation}
where $h$ and $k_\mathrm{B}$ are Planck constant and Boltzmann constant; $\bar{g_\mathrm{ff}}(\nu, T_\mathrm{e})$ is the velocity averaged Gaunt factor.  The integration of Equation~(\ref{equation:radio2X-ray}) at 2--10~keV and the observed X-ray flux suggests only $\sim0.01$~mJy at $\sim100$~GHz in an optically-thin FFE radio spectrum.  This predicted radio flux density is not sufficient at all to explain the observed HFE.  
This is the case of ionized gas even with a relatively high temperature, where X-ray can be produced efficiently.

Alternatively, at relatively low temperatures, 
hydrogen permitted lines are useful to probe free--free radio emission from ionized sources as the same origin. 
We consider homogeneous ionized broad-line region~(BLR) clouds with the diameter $d$, the electron temperature $T_\mathrm{e}$, and the electron density $n_\mathrm{e}$.  The free--free absorption~(FFA) process require an adequate optical thickness toward our line of sight to make a spectral peak at $>43$~GHz as is the case for the observed HFE spectrum; the peak frequency $\nu_\mathrm{p}$, where an optical depth $\tau$ becomes $\sim1$, divides optically-thick and optically-thin regimes at the lower and higher frequencies, respectively (Equation~\ref{equation:tau}).  The radio spectrum of a cloud shows an inverted shape with an index of $\alpha=+2$ and a nearly flat spectral shape with $\alpha=-0.1$ at the lower and higher regimes, respectively.  
Generally, possible physical parameters of a BLR cloud are $T_\mathrm{e} \sim 10^4$--$10^5$~K, $n_\mathrm{e} = 10^9$--$10^{12}$~cm$^{-3}$, and $d\sim10^{13}$~cm.   These parameter ranges make a spectral peak at $\nu_\mathrm{p} \gtrsim  140$~GHz, lower-end of which is roughly reasonable for the observed HFE component.  The flux density from a BLR cloud is expected to be $\sim1\times10^{-11}$~mJy at an optically-thin regime (Equation~\ref{equation:fluxdensity}), requiring the number of clouds $\sim3\times 10^{11}$ to achieve the observed radio emission of $\sim3$~mJy.  
In this context, the total mass of the clouds would be equivalent to $3\times10^2$--$3\times10^5\ M_\odot$.  The BLR size of NGC~985 with a broad-line luminosity of $\sim9.0\times10^{43}$~ergs~s$^{-1}$ \citep{Mullaney:2008} is expected to be $7.4\times10^{-2}$~pc from the nucleus, based on the relation between the broad-line luminosity and the reverberation measurement \citep{Kaspi:2005}.  For $\sim3\times10^{11}$~clouds with $d\sim10^{13}$~cm each, a (homogeneous) volume filling factor would be $3\times10^{-3}$, which is too large compared to a typical estimation of $\sim10^{-6}$ \citep[e.g., ][]{Netzer:1990}.  

On the other hand, narrow-line components in the narrow-line region or star forming regions also cannot be responsible for the HFE because their densities are too low to make opacity sufficiently high for $\nu_\mathrm{p}>43.3$~GHz.

\subsection{Free--free absorption: possible for HFE}\label{section:FFA}
This subsection provides physically possible parameters for a free--free absorber toward a putative background source, such as synchrotron jets intrinsically showing a steep spectrum; the FFA makes jet's spectrum convex by cutting off of the radio emission at frequencies lower than $\sim\nu_\mathrm{p}$.  Therefore, its low-frequency tail in the optically-thick regime would be observed as an HFE.  
We consider a small jet that is entirely confined within an ionized region, which is continuously photo-ionized up to the St\"{o}mgren radius from a central engine, as a very rough estimate.  

Eliminating $n_\mathrm{e}$ from Equations~(\ref{equation:tau}) and (\ref{equation:Stromgren}), the resultant equation gives $d$ depending on the optical depth $\tau_\nu$ at the frequency $\nu$, the ionizing photon rate $N_*$, and the electron temperature $T_\mathrm{e}$.  The observed HFE requires $\tau=1$ at $\nu>43.3$~GHz.  Assuming a fraction of luminosity at $>13.6$~eV (the Lyman edge) would contribute to keep a circumnuclear torus ionized, we estimate $N_*<1.3 \times 10^{53}$~photon~sec$^{-1}$ from the observed X-ray emission (Section~\ref{section:FFE}; \citealt{Krongold:2005}).  We assume $T_\mathrm{e}>10^4$~K.  The resultant diameter is $d<0.8$~pc for the ionized region.  Under this condition, an electron density is expected to be $n_\mathrm{e}>1.3 \times 10^5$~cm$^{-3}$ from Equation~(\ref{equation:tau}).   These physical conditions of the ionized region are consistent with those of plasma tori that were previously discovered in NGC~1068 \citep{Roy:1998,Gallimore:1997,Krips:2006} and NGC~1052 \citep{Kameno:2001,Sawada-Satoh:2008} with a physical scale of sub-pc around their central engines.  
On the other hand, such a free--free absorber itself cannot get to be a significant free--free radio emitter.  Eliminating $d$ from Equation~(\ref{equation:fluxdensity}) subsequently, the resultant equation indicates an upper limit of $S_\nu \ll 1$~mJy under the condition of $\tau_\nu=1$ at $\nu>43$~GHz, $N_*<1.3 \times 10^{53}$~photon~sec$^{-1}$, and $T_\mathrm{e} \ll 1.6\times10^8$~K ($70$~keV; see Section~\ref{section:FFE}).  Hence, such free--free plasma would not over-shine the background jet.

\subsection{Synchrotron self-absorption: unlikely for HFE?}\label{section:SSA}
For the case of a synchrotron jet, its self-absorbed radio emission potentially contributes the millimeter excess if the jet is still very compact due to being embedded in very dense environment at the nuclear region of this undergoing merger galaxy.  
Synchrotron self-absorption makes a power-law spectrum convex by absorption at lower  frequencies.    
Such a peaked radio spectrum also invokes Gigaheltz-Peaked Spectrum~(GPS) and Compact Steep Spectrum~(CSS) radio sources as progenitor candidates of radio galaxies \citep{ODea:1998}.        
GPS sources generally show spectral peaks at $\sim1$~GHz; sources showing spectral peaks at $>22$~GHz are very rare \citep{Dallacasa:2000}.
With the GPS/CSS samples, correlations among their peak frequency, peak flux density, and angular size provide strong evidence that synchrotron self-absorption is the cause of the spectral turnovers; the ratio of equipartition component size to observed overall linear size remains constant, $\sim5$--6, implying a self-similar evolution \citep{Snellen:2000}.  According to this correlation, on the putative HFE component of NGC~985, an overall linear size is expected to be $\sim0.01$~pc ($\sim0.002$~pc as the equipartition component size), corresponding to only $500\times$ Schwarzschild radius of the black hole with a mass of $\sim2.2\times10^8 M_\odot$ \citep{Kim:2008,Vasudevan:2009}.  
Indeed, on the basis of synchrotron theory, a synchrotron blob with $\sim3$~mJy at a peak frequency of $\sim100$~GHz as the HFE component of NGC~985 would be $\sim0.0017$~pc ($\approx$ 1/6-time the expected overall size) in radius and a magnetic field of 4.4~Gauss under the equipartition condition, according to \citet{Kino:2014} (eq.~16 and 18).  
Note that the physical scale of the case of NGC~985 is $\sim10^{2}$-times smaller than the established range of the empirical correlation throughout the GPS/CSS samples.  Such a small-sized blob would be short-lived: only $\sim0.6$~year, assuming an adiabatic expansion velocity of $0.1c$ or only 0.03~year in terms of synchrotron cooling in radiating electrons.  On the contrary, the HFE has been observed during 1985--2003~(Table~\ref{table:observation}); the HFE component has just declined into halves in the two decades (at 14.9~GHz; Figure~\ref{figure:spectrum}.  Thus, the predicted life times for such a compact synchrotron jet are completely different from the observations.  Intermittently recurrent activity is required to explain that the HFE apparently survives, under the scenario of synchrotron self-absorption.

\subsection{Synchrotron from accretion flow: insufficient for HFE}\label{section:ADAF}

Both X-ray emission with a cut-off at $\sim100$~keV and an inverted radio spectrum can be produced by hot accretion flow, such as the RIAF.  We consider the model of the advection-dominated accretion flow~\citep[ADAF;][]{Narayan:1994}, which is a model taking into account thermal synchrotron process in radio and can explain many SEDs of low-luminosity AGNs~(LLAGNs) with low accretion rates \citep{Eracleous:2010,Nemmen:2014}.  On the other hand, many observations for nearby LLAGNs indicate that radio luminosities roughly an order of magnitude greater than those predicted by the ADAF model \citep{Ulvestad:2001, Anderson:2004, Doi:2005, Doi:2011}; more efficient (thermal) synchrotron emission from a jet base, which may corresponds to a steady jet in X-ray binaries in the low/hard state, may additionally contribute the observed radio luminosities \citep{Falcke:1999, Yuan:2002a,Markoff:2008}.  The LLAGN NGC~266 is an example clearly showing the coexistence of steep and inverted spectral components \citep{Doi:2005a}, the latter of which can be attributed to the ADAF and jet base components.  

However, NGC~985 is a relatively luminous Seyfert with a Eddington ratio of 0.02, which is possibly higher than a critical accretion rate for the application of the ADAF model.  We attempt to calculate an expected radio flux density from the ADAF according to the model presented by \citet{Mahadevan:1997}; as a result, only $\sim0.04$~mJy would be expected at 43.3--95.7~GHz from thermal hot electrons with $\sim1 \times 10^9$~K; even a ten-times larger value as a jet base would be insufficient for the observed HFE.  This discussion suggests only thermal synchrotron process cannot be responsible for the observed luminosity.

\subsection{Synchrotron from the standard disk corona: sufficient for HFE?}\label{section:corona}
\citet{Inoue:2014} presented that the synchrotron from nonthermal electrons in corona above an accretion disk of nearby Seyfert galaxies is potentially detectable at the submillimeter regime using the ALMA sensitivities.  A strong correlation is known to exist between the quiescent radio and X-ray emission in coronally active cool stars: radio-to-X-ray ratio $L_\mathrm{R}/L_\mathrm{X} \simeq 10^{-5}$ \citep{Guedel:1993}.  If the same mechanism magnetically heats corona above accretion disks, AGNs may also satisfy this relation; \citet{Laor:2008} predicts that the signature of X-ray emitting corona from radio-quiet quasars possibly appears in the millimeter band, because of synchrotron self-absorption making spectra strongly inverted with the index $\alpha\leqq+2.5$, which is significantly flatter than that of dust spectrum ($\alpha\sim+3$--$4$) in the Rayleigh-Jeans region (Section~\ref{section:dust}).  
For NGC~985\footnote{We use an absorption-corrected X-ray luminosity at $0.2$--$20$~keV, which was converted from that at $2$--$10$~keV with the photon index $\Gamma=1.96$ \citep{Parker:2014}, and an radio luminosity at 5~GHz estimated by a spectral extrapolation with $\alpha=-0.1$ (optically-thin) from the observed flux density at 43.3~GHz for the HFE component.}, $L_\mathrm{R}/L_\mathrm{X} = 0.4 \times 10^{-5}$, which is consistent with this correlation.  

Using the model of \citet{Inoue:2014}, we found realistic parameters of the physical condition of corona that is required from both the observed luminosities of X-ray and HFE radio emissions from the NGC~985 nucleus: $100\times$ Schwarzschild radius for the corona size, an equipartition magnetic field of 150~Gauss, the 10\%-Eddington ratio, an electron temperature of 100~keV, and the non-thermal energy fraction $\sim4$\%.   Such a corona size is roughly consistent with the sizes estimated by observations for several AGNs possibly showing millimeter excesses \citep{Behar:2015} and the day-scale variability \citep{Baldi:2015}.  At this time, our radio data are still in a limited quality.  Precise SED modeling based on (quasi-)simultaneous ALMA--X-ray observations would be demanded to discriminate the coronal origin from other processes.

\section{Summary and conclusion}\label{section:conclusion} 
The radio spectrum including high-frequency excess~(HFE) that previously discovered by \citet{Antonucci:1988,Barvainis:1996} toward the Seyfert nucleus in NGC~985 has been confirmed by our new observations at higher radio frequencies as well.  The steep spectrum at low frequencies can be interpreted as the stellar origin of the star-forming activity in this merging galaxy.  Dust emission at the Rayleigh-Jeans regime, compact jets under synchrotron self-absorption, or thermal synchrotron from hot accretion flow such as ADAF cannot be responsible for the observed HFE.  On the other hand, we cannot rule out the free--free emission from BLR clouds, synchrotron jets free--free absorbed by a circumnucler photo-ionized region, and self-absorbed nonthermal synchrotron from disk corona as the origin of the HFE, with limited qualities of data at this time.  

Conclusive evidence for determining the origin of HFE would be obtained by more detailed high-frequency radio spectrum from future ALMA observations with a high-angular resolution (sufficient for resolving extended components) in cooperation with (quasi)-simultaneous observations at optical (BLR clouds radiating free--free emission) and X-ray (corona radiating nonthermal synchrotron) to investigate the correlation of time variations.

\begin{ack}
We used the US National Aeronautics and Space Administration's (NASA) Astrophysics Data System~(ADS) abstract service and NASA/IPAC Extragalactic Database (NED), which is operated by the Jet Propulsion Laboratory~(JPL).  
The National Radio Astronomy Observatory is a facility of the National Science Foundation operated under cooperative agreement by Associated Universities, Inc. 
We acknowledge the support of the Nobeyama Radio Observatory~(NRO) staff in operating the telescopes and for their continuous efforts to improve the performance of the instruments.  The NRO is a branch of the National Astronomical Observatory of Japan~(NAOJ), which belongs to National Institutes of Natural Sciences~(NINS).   
This study was partially supported by Grants-in-Aid for Scientific Research (B; 24340042, AD) from the Japan Society for the Promotion of Science (JSPS).   
\end{ack}


\appendix
\section{A simple free--free plasma model}\label{appendix}
We prepare formula for a simple model to evaluate the observed SED of NGC~985 by free--free emission/absorption.  We consider a spherical fully-ionized region with the diameter $d$ and homogeneous profiles of the electron density $n_\mathrm{e}$ and the temperature $T_\mathrm{e}$.   
In the range of Gaunt factor in radio wavelengths, the optical depth $\tau_\nu$ at the frequency $\nu$ is approximated by the following formula in astronomically convenient units \citep{Mezger:1967}:  
\begin{equation}\label{equation:tau}
\tau_\nu \approx 0.08235 \left( \frac{T_\mathrm{e}}{\mathrm{K}} \right)^{-1.35} \left( \frac{\nu}{\mathrm{GHz}} \right)^{-2.1}
\left( \frac{EM}{\mathrm{pc}\ \mathrm{cm}^{-6}} \right),   
\end{equation}
where $EM$ is the emission measure as 
\begin{eqnarray*}
\frac{EM}{\mathrm{pc}\ \mathrm{cm}^{-6}} &\equiv&
\int_\mathrm{los} \left( \frac{n_\mathrm{e}}{\mathrm{cm}^{-3}} \right)^2
d\left( \frac{s}{\mathrm{pc}} \right) \\ \nonumber  
&=& \frac{2}{3} \left( \frac{n_\mathrm{e}}{\mathrm{cm}^{-3}} \right)^2 \left( \frac{d}{\mathrm{pc}} \right) \nonumber, 
\end{eqnarray*}
where $s$ is the path length through ionized gas.  The factor 2/3 is attributed to the spherical geometry.   

The observed flux density at the frequency $\nu$ of free--free emission is    
\begin{eqnarray}\label{equation:fluxdensity}
S_\nu &=& \frac{2 k_\mathrm{B} T_\mathrm{B} \nu^2}{c^2} \Omega \\ \nonumber
      &=& 2.408 \times 10^4 \left( \frac{T_\mathrm{e}(1-e^{-\tau_\nu})}{\mathrm{K}} \right)  \left( \frac{\nu}{\mathrm{GHz}} \right)^{2}  \left( \frac{d}{D} \right)^2  \ \ \mathrm{in\ Jy}, 
\end{eqnarray}
where $k_\mathrm{B}$ is Boltzmann constant, $T_\mathrm{B}$ is the brightness temperature, $c$ is the speed of light, $\Omega$ is the solid angle of an emitting source, and $D$ is the distance to the source.  

An ionizing photon rate, $N_*$, for a region that is continuously photo-ionized up to the St\"{o}mgren radius from a central engine can be related to by the following equation: 
\begin{eqnarray}\label{equation:Stromgren}
N_* &=& \frac{4}{3}\pi \left( \frac{d}{2} \right)^3 n_\mathrm{e}^2 \alpha_\mathrm{B} \\ \nonumber
    &\approx& 4.0 \times 10^{46} \left( \frac{d}{\mathrm{pc}} \right)^3     \left( \frac{n_\mathrm{e}}{\mathrm{cm}^{-3}} \right)^2    \left( \frac{T_\mathrm{e}}{\mathrm{K}} \right)^{-0.85} \ \ \mathrm{in\ photon\ s^{{-1}}}, 
\end{eqnarray}
where $\alpha_\mathrm{B} \approx 2.6\times10^{-13} (T_\mathrm{e}/10^4\mathrm{K})^{-0.85}$~cm$^3$~s$^{-1}$ is the hydrogen recombination coefficient.

\end{document}